\documentstyle[epsf,cite,aps,amssymb]{revtex}

\newcommand{\be}{\begin{equation}}
\newcommand{\ee}{\end{equation}}
\newcommand{\bea}{\begin{eqnarray}}
\newcommand{\eea}{\end{eqnarray}}
\newcommand{\ba}{\begin{array}}
\newcommand{\ea}{\end{array}}
\newcommand{\nn}{\nonumber \\}
\newcommand{\half}{\frac{1}{2}}

\newcommand{\Z}{\mathbb{Z}}
\newcommand{\R}{\mathbb{R}}

\begin{document}

\title{Off-Shell Duality in Maxwell and Born-Infeld Theories}

\author{Victor O. Rivelles}

\address{Instituto de F\'\i sica, Universidade de S\~ao Paulo\\
 Caixa Postal 66318, 05315-970, S\~ao Paulo - SP, Brazil\\
E-mail: rivelles@fma.if.usp.br}

\maketitle

\begin{abstract}
It is well known that the classical equations of motion of Maxwell and
Born-Infeld theories are invariant under a duality symmetry acting on
the field 
strengths. We review the implementation of the $SL(2,\Z)$ duality in these
theories as linear but non-local transformations of the
potentials \cite{us}. 

\end{abstract}
 
\vspace{ 1cm}

In field theory duality is often realized as symmetries of the
equations of motion. It is known that the equations of motion,
Bianchi identities and the energy-momentum tensor of Maxwell and
Born-Infeld theories are invariant under $SO(2)$  
rotations which mix the electric and magnetic fields
\cite{Gibbons-Rasheed,Gaillard-Zumino}. The action, 
however, is not invariant. The $SO(2)$ symmetry
can be enlarged to $SL(2,\R)$ when a dilaton and an axion are added
\cite{Gibbons-Rasheed2, Gaillard-Zumino}. 
However, it is
desirable that such symmetries could be implemented at the quantum
level as symmetries of the action and partition function.
In this way the symmetries will hold in any situation and not only for
on-shell quantities.  Off-shell
symmetries must be implemented in the 
basic field variables (and not in the field strengths for gauge
theories) either in the Lagrangian or Hamiltonian
formalism. When varying the action the resulting boundary term must be
local in time, giving rise to a Noether current associated to the
invariance. However, the boundary term can be non-local in space
provided that it has a sufficient falloff at spatial infinity. 
This allows  the variations of the basic field variables to be
non-local in space. These ideas were
first explored in \cite{Deser-Teitelboim} where the $SO(2)$ symmetry of
Maxwell equations were implemented at the 
action level in the Hamiltonian formalism in Coulomb gauge. 
The same holds
for the Born-Infeld theory \cite{Bengtsson} and for gauge theories
coupled to matter and gravity \cite{Igarashi}. 
These transformations leave the action invariant only in
the Coulomb gauge. This could be seen as a drawback since the symmetry
manifests itself only in a particular gauge. Even so, it may be quite
useful. A typical example is the Chern-Simons theory in Landau
gauge. In this case there appears a vector supersymmetry
\cite{Chern-Simons} which can be extended to the exceptional algebra
$D(2,\alpha)$ \cite{Damgaard}. This symmetry is essential to show the
renormalizability of the model. 

We should point out that there is an alternative procedure to
implement off-shell symmetries in the action with local transformation
laws. Usually it breaks manifest Lorentz invariance and demands the
introduction 
of more fields. For the Maxwell theory this requires a description in
terms of two potentials giving rise to the Schwarz-Sen model
\cite{Schwarz-Sen} or, alternatively, an infinite number of them 
\cite{McClain}. Duality manifests itself as rotations between the
potentials. It is possible to show \cite{Girotti-Gomes-Rivelles} that
the duality symmetry of the 
Schwarz-Sen model is the local form of the non-local transformations
found in \cite{Deser-Teitelboim}. Although the Schwarz-Sen 
model is not manifestly Lorentz covariant this symmetry can be made
manifest by the inclusion of auxiliary fields and some gauge symmetry
through 
the PST formalism \cite{PST}. A similar situation is found for the
Born-Infeld theory \cite{Parra}. It should be remarked that
this situation is 
not exclusive of duality symmetry. Even well known symmetries, like the
BRST symmetry, can be cast into a non-local form at the expense of
loosing manifest Lorentz invariance \cite{Rivelles}. 

The $SL(2,\R)$ symmetry of the equations of motion found when a
dilaton and an axion are added, manifests, at the quantum level, as an
$SL(2,\Z)$ duality of the partition function. This happens when the
dilaton and the axion take their vacuum expectation value which are
combined into a complex coupling constant $\tau$ with its real part
being the theta term. Now the action and the partition function are
functions of $\tau$ and duality manifests as modular transformations
of the coupling constant. In Maxwell theory the Lagrangian partition
function is found to be a 
modular form under $SL(2,\Z)$ transformations of the coupling
constant $\tau$ \cite{Witten,Olive-Alvarez}. At the
Hamiltonian level, the partition function is modular invariant with
modular weight equal to zero \cite{Lozano,Olive-Alvarez}. In this case
duality can be implemented as a canonical transformation on the
reduced phase space. This means that Gauss law holds and we are
on-shell. Recently this duality was implemented off-shell as linear but
non-local transformations of the potentials as we will describe below
\cite{us}. 

Maxwell theory with a theta term is described by the following action
in Minkowski space with metric $(+---)$ 
\be
\label{2.1}
S = - \frac{1}{8 \pi} \int d^4x \,\, \left( \frac{4 \pi}{g^2} F^{\mu\nu}
F_{\mu\nu} + \frac{\theta}{2\pi} F^{\mu\nu} {}^{*}F_{\mu\nu} \right),
\ee
where ${}^{*}F^{\mu\nu} = \frac{1}{2} \epsilon^{\mu\nu\rho\sigma}
F_{\rho\sigma}$. The Hamiltonian formulation is obtained as usual. 
There is a primary constraint $\Pi^0 =
\frac{\delta L}{\delta \dot{A}_0} = 0$ and the secondary constraint
receives no contribution from the theta term, giving rise to the usual
Gauss law $\partial_i \Pi^i=0$. The Hamiltonian density is then  
\be
\label{2.2}
H_M = - \frac{2 \pi i}{\tau - \overline{\tau}} \Pi^i \Pi_i - i
\frac{\tau + \overline{\tau}}{\tau - \overline{\tau}} \Pi^i B_i -
\frac{i}{2\pi} \frac{\tau \overline{\tau}}{\tau - \overline{\tau}} B^i
B_i,
\ee
where $\Pi^i = \frac{\delta L}{\delta \dot{A}_i}$, the magnetic field
is $B_i = \frac{1}{2} \epsilon_{ijk}F^{jk}$ 
and the complex coupling constant is $\tau=\frac{\theta}{2\pi} +
\frac{4\pi i}{g^2}$. The contribution from the theta term appears in the
second and third terms of Eq.(\ref{2.2}). 

The BRST charge is the same as in pure Maxwell theory since the
constraint structure was not modified 
\be
\label{3}
Q = \int d^3x \,\, \left( \partial_i \Pi^i C + {\cal P}_D \Pi_0 \right).
\ee
The ghosts obey the canonical Poisson brackets $\{ {\cal P}_C, C
\} = \{ {\cal P}_D, D \} = -1$ and the BRST transformations
are
\bea
\label{4}
&& \delta A_i = \partial_i C,  \qquad \delta A_0 = - {\cal P}_D, \qquad
\delta \Pi_0 = 0,  \qquad \delta \Pi_i = 0, \nn
&& \delta C= 0, \qquad\quad\,  \delta {\cal P}_C = - \partial_i \Pi^i,
\quad\,\,\, \delta D = - \Pi_0, \,\,\,\,\, \delta {\cal P}_D = 0. 
\eea
The partition function is then
\be
\label{5}
Z(\tau) = \int {\cal D}A_\mu \,\, {\cal D}\Pi_\nu \,\, {\cal
  D}(\mbox{ghosts}) \,\, e^{-i S_M(\tau)}, 
\ee
where the Maxwell effective action is 
\be
\label{6}
S_M(\tau) = \int d^4x \,\, \left( \Pi^\mu \dot{A}_\mu + \dot{C} {\cal
  P}_C   + \dot{\cal P}_D D  - H_M - \{ Q, \Psi \}   \right),
\ee
and $\Psi$ is the gauge fixing function. 

In order to consider the $SL(2,\Z)$ duality it is convenient
to split the vector fields $A_i$ and $\Pi_i$ into their transversal
$A^T_i, \Pi^T_i$ and longitudinal parts $A^L_i, \Pi^L_i$. We will also
consider finite $SL(2,\Z)$ transformations. We have found that the
$SL(2,\Z)$ transformations are given by \cite{us}
\bea
\label{3.1}
&A^T_i& = a \tilde{A}^T_i + 2\pi c \epsilon_{ijk}
\frac{\partial^j}{\partial^2} \tilde{\Pi}^{Tk}, \qquad  \Pi^T_i = d
\tilde{\Pi}^T_i + \frac{b}{2\pi} \tilde{B}_i, \nn
&A^L_i& = |a-c \tilde{\tau}| \tilde{A}^L_i, \qquad\qquad\qquad\,\,  \Pi^L_i =
\frac{1}{|a-c \tilde{\tau}|} \tilde{\Pi}^L_i, \nn
&A_0& = |a-c \tilde{\tau}| \tilde{A}^0, \qquad\qquad\qquad\,\,\,  \Pi_0 =
\frac{1}{|a-c   \tilde{\tau}|} \tilde{\Pi}_0, \nn
&C&= |a-c \tilde{\tau}| \tilde{C}, \qquad\qquad\qquad\,\,\,\,\, {\cal P}_C =
\frac{1}{|a-c \tilde{\tau}|} \tilde{{\cal P}}_C, \nn
&D& = \frac{1}{|a-c \tilde{\tau}|} \tilde{D},
\qquad\qquad\qquad\,\,\,\, {\cal P}_D = |a-c \tilde{\tau}| \tilde{\cal P}_D, 
\eea
\be
\label{3.2}
\tilde{\tau} = \frac{a \tau + b}{c \tau + d},
\ee
where $a,b,c$ and $d$ are integers satisfying $ad-bc=1$. 
Notice that the transversal (or physical) part of the
vectors are transformed among themselves while the gauge dependent (or
non-physical) parts, which include: $A_0, A_i^L,\Pi_0$ and $\Pi^L_i$
and the ghosts, transform into themselves.  
These transformations are local in time but non-local in space. 

The Hamiltonian density Eq.(\ref{2.2}) is not invariant under
Eqs.(\ref{3.1},\ref{3.2}). It transforms 
as
\bea
\label{3.3}
H_M =  \tilde{H}_M &-& \frac{2\pi i}{\tilde{\tau} - \overline{\tilde{\tau}}}
\frac{2(a - |a-c\tilde{\tau}|) - c(\tilde{\tau} +
  \overline{\tilde{\tau}})}{|a-c\tilde{\tau}|} \tilde{\Pi}^{Ti}
\tilde{\Pi}^L_i \nn
&-& \frac{i}{\tilde{\tau} - \overline{\tilde{\tau}}}
\frac{(a - |a-c\tilde{\tau}|)(\tilde{\tau} + \overline{\tilde{\tau}}) - 2c
  \tilde{\tau}\overline{\tilde{\tau}}}{|a-c\tilde{\tau}|}
\tilde{\Pi}^{Li} \tilde{B}_i,
\eea
and upon integration the extra terms give rise to surface
contributions. The kinetic terms in the effective action Eq.(\ref{6})
are also invariant up to surface terms. Hence, the Hamiltonian is
modular invariant up to surface terms. 

The gauge fixing term in Eq.(\ref{6}) requires some
care but it can be proved to be BRST invariant \cite{us}.
We then conclude that the effective action is modular invariant for
any gauge  
choice. The Jacobian of the transformations Eqs.(\ref{3.1}) can be
computed and it is found to be equal to one. Therefore, they can be 
regarded as a canonical transformation. Hence the path integral
measure is also invariant. As a consequence, the partition function is
modular invariant. It should be stressed that
the partition function in the Lagrangian formalism is not modular
invariant under duality, rather it transforms as a
modular form \cite{Witten}. However, the phase space partition
function is modular invariant \cite{Lozano,Olive-Alvarez}. 
It can also be shown that Eqs.(\ref{3.1}) reduce to the
familiar duality transformations of the classical equations of
motion. 

It is well known that the Born-Infeld theory has an $SO(2)$ symmetry in
its classical equations of motion which can be extended to $SL(2,\R)$
if an axion and a dilaton are added \cite{Gibbons-Rasheed2}. If we
consider just the axion and dilaton vacuum expectation values we
get a Born-Infeld theory with a theta term.  The action is 
\be
\label{4.1}
S = \int d^4x \,\, \left( 1 -\frac{\theta}{16 \pi^2}
F^{\mu\nu}{}^*F_{\mu\nu} - \sqrt{1 + \frac{1}{g^2} F^{\mu\nu}
F_{\mu\nu} - \frac{1}{4 g^4} ( F^{\mu\nu} {}^*F_{\mu\nu} )^2 }
\right). 
\ee
In the weak field limit it reduces to Maxwell theory with a theta term
Eq.(\ref{2.1}). There is also a dimensionful constant in the action
which was set equal to one. In order to handle the square root in the
action we introduce an auxiliary field $V$
\be
\label{4.2}
S = \int d^4x \,\, \left[ 1 -\frac{\theta}{16 \pi^2} F^{\mu\nu}
{}^*F_{\mu\nu} - \frac{V}{2} \left( 1 + \frac{1}{g^2} F^{\mu\nu}
F_{\mu\nu} - \frac{1}{4 g^4} ( F^{\mu\nu} {}^*F_{\mu\nu} )^2 \right) -
\frac{1}{2V} \right].  
\ee

The Hamiltonian formulation is straightforward and follows closely
that of \cite{Parra}. Since we have introduced an auxiliary field
$V$ there are two primary constraints $\Pi_0=0$ and
$p=\frac{\partial L}{\partial \dot{V}}=0$. From the first constraint
we get as secondary constraint the Gauss law. From the second
constraint we get an algebraic equation for $V$ which can be solved
so that $V$ is eliminated. We then find the Hamiltonian density 
\be
\label{4.3}
H_{BI} = \sqrt{1 + 2 H_M - (\Pi^i  B_i)^2 + B^i B_i \Pi^j \Pi_j } - 1.
\ee
Clearly the Hamiltonian is not modular invariant under
Eqs.(\ref{3.1},\ref{3.2}). The Maxwell Hamiltonian 
is not invariant and since the duality transformations are linear
the extra terms  in Eq.(\ref{3.3}) can not be
canceled against those coming from $(\Pi B)^2 - B^2 \Pi^2$ term in the
square root in Eq.(\ref{4.3}). Either non-linear terms must be
introduced in Eqs.(\ref{3.1}) or something else must be modified. 

It must be noted that both the Maxwell and the Born-Infeld
Hamiltonian densities can be  rewritten in terms of a complex vector field 
\be
\label{4.4}
P_i = \Pi_i + \frac{\tau}{2\pi} B_i.
\ee
We find that 
\be
\label{4.5}
H_M = - \frac{2\pi i}{\tau - \overline{\tau}} P^i \overline{P}_i,
\ee
and
\be
\label{4.6}
H_{BI} = \sqrt{1 - \frac{4\pi i}{\tau - \overline{\tau}} P^i
\overline{P}_i - \frac{4 \pi^2}{(\tau - \overline{\tau})^2} (P
\times \overline{P})^2 } -1,
\ee
where the overline denotes complex conjugation. The vector $P_i$
transforms under duality as 
\be
\label{4.7}
P_i = \frac{1}{a - c \tilde{\tau}} \left(\tilde{\Pi}^T_i +
\frac{a-c\tilde{\tau}}{|a-c\tilde{\tau}|} \tilde{\Pi}^L_i +
\frac{\tilde{\tau}}{2\pi} \tilde{B}_i \right),
\ee
while 
\be
\frac{1}{\tau - \overline{\tau}} = \frac{ |a - c
  \tilde{\tau} |^2 }{\tilde{\tau} - \overline{\tilde{\tau}}}.
\ee
This explains why the Maxwell Hamiltonian is not invariant. The
longitudinal and transversal parts of $\tilde{\Pi}_i$ do not combine
themselves back into $\tilde{\Pi}_i$ so that $P_i$ is not a modular
form. If instead of $|a-c\tilde{\tau}|$ in the denominator of the
$\tilde{\Pi}^L_i$ term in Eq.(\ref{4.7}) we had just $a-c\tilde{\tau}$
we could recover $\tilde{P}_i$. But taking out the modulus in the
transformations Eqs.(\ref{3.1}) is not consistent because all fields
are real. On the other side if we could change only the transformation
for $\Pi^L_i$ that would do the job. It is then necessary that
$\Pi^L_i$ possess an imaginary part. For consistency $A_0, \Pi_0,
A_i^L$ and the ghosts must have an imaginary part as well.

So we start with the non-physical sector $A_0, \Pi_0, A^L_i, \Pi^L_i$
and the ghosts all described by complex fields. Since the number of ghosts
has also doubled the number of physical degrees of freedom is still
the same. The vectors $A_i$ and $\Pi_i$ are now complex with their
transversal part taken to be real while their longitudinal parts are
taken to be complex. The effective action is now
\be
\label{4.8}
S_{BI} = \int d^4x \,\, \left( \half \overline{\Pi}^\mu \dot{A}_\mu + \half
\Pi^\mu \dot{\overline{A}}_\mu + \half
\dot{C} \overline{\cal P}_C + \half \dot{\overline{C}} {\cal P}_C +
\half \dot{\cal P}_D \overline{D} + \half \dot{\overline{\cal P}}_D
D - H_{BI} - \{Q,\Psi\} \right),
\ee
The Hamiltonian density has the same form as in Eq.(\ref{4.6}) with
$P_i$ defined by Eq.(\ref{4.4}) but with complex fields instead of
real fields. The integrand in the square root in Eq.(\ref{4.6}) is real. 

The BRST charge is now
\be
\label{4.9} 
Q = \half \int d^3x \,\, \left( \partial_i \Pi^i \overline{C} +
\partial_i \overline{\Pi}^i C + \overline{\cal P}_D  \Pi_0 +
{\cal P}_D \overline{\Pi}_0 \right),
\ee
so that $Q$ is real. The BRST transformations are modified in a
straightforward way. The gauge fixing fermion reads now
\be
\label{4.10}
\Psi = \half \int d^3x \,\, \left( \chi \overline{D} + \overline{\chi}
D + A_0 \overline{\cal P}_C + \overline{A}_0 {\cal P}_C \right), 
\ee
and is also real. It can be shown that this theory is equivalent to
the original Born-Infeld theory \cite{us}. 

Then the $SL(2,\Z)$ duality transformations are now 
\bea
\label{4.12}
&A^T_i& = a \tilde{A}^T_i + 2\pi c \epsilon_{ijk}
\frac{\partial^j}{\partial^2} \tilde{\Pi}^{Tk}, \qquad  \Pi^T_i = d
\tilde{\Pi}^T_i + \frac{b}{2\pi} \tilde{B}_i, \nn
&A^L_i& = (a-c \overline{\tilde{\tau}}) \tilde{A}^L_i,
\qquad\qquad\qquad\,  \Pi^L_i = 
\frac{1}{a-c \tilde{\tau}} \tilde{\Pi}^L_i, \nn
&A_0& = (a-c \overline{\tilde{\tau}}) \tilde{A}^0, \qquad\qquad\qquad\,\,
\Pi_0 = \frac{1}{a-c   \tilde{\tau}} \tilde{\Pi}_0, \nn
&C&= (a-c \overline{\tilde{\tau}}) \tilde{C}, \qquad\qquad\qquad\,\,\,\,
{\cal P}_C =
\frac{1}{a-c \tilde{\tau}} \tilde{{\cal P}}_C, \nn
&D& = \frac{1}{a-c \tilde{\tau}} \tilde{D},
\qquad\qquad\qquad\quad\, {\cal P}_D = (a-c \overline{\tilde{\tau}})
\tilde{\cal P}_D.
\eea
The unphysical sector is composed of modular forms. The vector $P_i$
is also a modular form. It transforms as 
\be
\label{4.13}
P_i = \frac{1}{a-c\tilde{\tau}} \tilde{P}_i,
\ee
so that the Maxwell Hamiltonian is modular invariant with no surface
terms being generated. The Born-Infeld Hamiltonian is also modular
invariant. It is easy to show that 
the kinetic terms in the effective action Eq.(\ref{4.8}) are also
invariant up to surface terms. The BRST charge Eq.(\ref{4.9})
is also invariant. It can also be shown that the gauge fixing term in
Eq.(\ref{4.8}) is 
also modular invariant so that the effective action Eq.(\ref{4.8}) is
modular invariant. Finally we can show that the duality transformations
Eqs.(\ref{4.12}) have a unity Jacobian so that the partition function is
modular invariant. Also, the duality transformations Eqs.(\ref{4.12})
reduce to the usual 
duality transformations of the classical equations of motion. 

We have shown how it is possible to generalize the $SL(2,\R)$ symmetry
of the equations of motion, for Maxwell and Born-Infeld theories, to an
off-shell duality.  For the Maxwell theory we found that the
Hamiltonian $H_M$ is modular invariant up to a surface term. 
In the Born-Infeld case it was necessary to
consider the longitudinal part of the fields as complex fields. Then
the Born-Infeld Hamiltonian $H_{BI}$ is strictly modular 
invariant with no boundary terms being generated by the
transformation. Of course, we could consider Maxwell theory with the
longitudinal part of the fields being complex as well. In this case
the Hamiltonian would be modular invariant without any boundary term. However
there is no clear interpretation for the complex
longitudinal fields introduced in these theories. 

\vspace{1cm}

This work is partially supported by FAPESP and PRONEX
No. 66.2002/1998-9. The work of V. O. Rivelles is partially supported 
by CNPq. 
\nopagebreak
\samepage


\begin{references}

\bibitem{us}C. R. Le\~ao and V. O. Rivelles, ``Off-Shell Duality in
  Born-Infeld Theory'', Nucl. Phys. {\bf B602} 514 (2001),
  hep-th/0101031. 
\bibitem{Gibbons-Rasheed} G. W. Gibbons and D. A Rasheed,
  ``Electric-Magnetic Duality Rotations in Non-Linear Electrodynamics'',
  Nucl. Phys. {\bf B454} (1995) 185, hep-th/9506035.
\bibitem{Gaillard-Zumino} M. K. Gaillard and B. Zumino, ``Duality
  Rotations for Interacting Fields'', Nucl. Phys. {\bf B193} (1981)
  221; 
``Self-Duality in Nonlinear Electromagnetism'',
  hep-th/9705226. 
\bibitem{Gibbons-Rasheed2} G. W. Gibbons and D. A Rasheed, ``$SL(2,R)$
  Invariance of Non-Linear Electrodynamics Coupled to an Axion and a
  Dilaton'', Phys.Lett. {\bf B365} (1996) 46, hep-th/9509141. 
\bibitem{Deser-Teitelboim} S. Deser and C. Teitelboim, ``Duality
  Transformations of Abelian and Non-Abelian Gauge Fields'', Phys
  Rev. {\bf D13} (1976) 1592.
\bibitem{Bengtsson} I. Bengtsson, ``Manifest Duality in Born-Infeld
  Theory'', Int. J. Mod. Phys. {\bf A12} (1997) 4869, hep-th/9612174. 
\bibitem{Igarashi} Y. Igarashi, K. Itoh and K. Kamimura,
  ``Electric-Magnetic Duality Rotations and Invariance of Actions'',
  Nucl.Phys. {\bf B536} (1998) 454, 
  hep-th/9806160;  
``Self-Duality in Super $D3$-brane Action'',
  Nucl.Phys. {\bf B536} (1998) 469, 
  hep-th/9806161. 
\bibitem{Chern-Simons} D. Birmingham, M. Rakowski and  G. Thompson,
  ``Renormalization of Topological Field Theory'', Nucl. Phys. {\bf
  B329} (1990) 83.
\bibitem{Damgaard} P. H. Damgaard and V. O. Rivelles,
  ``Symmetries of Chern-Simons Theory in Landau Gauge'', Phys. Lett. 
{\bf 245 B} (1990) 48. 
\bibitem{Schwarz-Sen} D. Zwanziger, ``Local Lagrangian Quantum Field
  Theory of Electric and Magnetic Charges'', Phys. Rev. {\bf D3}
  (1971) 880; 
J. H. Schwarz and A. Sen, ``Duality Symmetric
  Actions'', Nucl. Phys. {\bf B411} (1994) 35, hep-th/9304154.
\bibitem{McClain} B. McClain, Y. S. Wu and F. Yu, ``Covariant
  Quantization of Chiral Bosons and $OSP(1,1|2)$ Symmetry'',
  Nucl. Phys. {\bf B343} (1990) 689; 
I. Bengtsson and A. Kleppe, ``On
  Chiral p-Forms'', 
  Int. J. Mod. Phys. {\bf A12} (1997) 3397,  hep-th/9609102;
  N. Berkovits, ``Local Actions with Electric and Magnetic Sources'',
  Phys. Lett. {\bf B395} (1997) 28, hep-th/9610134.
\bibitem{Girotti-Gomes-Rivelles} H. O. Girotti, M. Gomes,
  V. O. Rivelles and A. J. Silva, ``Duality Symmetry in the
  Schwarz-Sen Model'', Phys. Rev. {\bf D56} (1997) 6615, hep-th/
  9702065.
\bibitem{PST} P. Pasti, D. Sorokin and M. Tonin, ``Duality Symmetric
  Actions with Manifest Space-Time Symmetries'', Phys.Rev. {\bf D52}
  (1995) 4277, hep-th/9506109.
\bibitem{Parra} D. Berman, ``$SL(2,\Z)$ Duality of Born-Infeld Theory
  from Nonlinear Selfdual Electrodynamics in Six-Dimensions'',
  Phys. Lett. {\bf B409} (1997) 153, hep-th/9706208; 
A. Khoudeir and
  Y. Parra, ``On Duality in the Born-Infeld Theory'', Phys. Rev. {\bf
  D58} (1998) 025010, hep-th/9708011. 
\bibitem{Rivelles} V. O. Rivelles, ``Comment on ``A New Symmetry for
  QED'' and  ``Relativistically Covariant Symmetry in QED'''',
  Phys. Rev. Lett. {\bf 75} (1995) 4150, hep-th/9509028; 
``Several
  Guises of the BRST Symmetry'', Phys. Rev. {\bf D53} (1996) 3247,
  hep-th/9510136.
\bibitem{Witten} E. Witten, ``On S-Duality in Abelian Gauge Theory'',
  hep-th/9505186; E. Verlinde, ``Global Aspects of Electric-Magnetic
  Duality'',  Nucl.Phys. {\bf B455} (1995) 211, hep-th/9506011. 
\bibitem{Olive-Alvarez} D. I. Olive and M. Alvarez, ``Spin and Abelian
  Electromagnetic Duality on Four-Manifolds'', hep-th/0003155.
\bibitem{Lozano} Y.Lozano, ``S-Duality in Gauge Theories as a
  Canonical Transformation'', Phys.Lett. {\bf B364} (1995) 19,
  hep-th/9508021; 
``Duality and Canonical Transformations'',
  Mod. Phys. Lett. {\bf A11} (1996) 2893, hep-th/9610024. 
\end{references}
\end{document}